 \documentclass{aastex}

\def\jref#1 #2 #3 #4 {{\par\noindent \hangindent=2em \hangafter=1
      \advance \rightskip by 0em #1, {\it#2}, {\bf#3}, #4.\par}}
\def\rref#1{{\par\noindent \hangindent=2em \hangafter=1
      \advance \rightskip by 0em #1.\par}}

\def\lsim{\,\lower2truept\hbox{${< \atop\hbox{\raise4truept\hbox{$\sim$}}}$}\,}
\def\gsim{\,\lower2truept\hbox{${> \atop\hbox{\raise4truept\hbox{$\sim$}}}$}\,}

\def\etal   {et~al.\,}

\slugcomment{Submitted to ApJ and received, 23 January 2001}

\shorttitle{Cosmological parameter determination from {\sc Planck} and SDSS}
\shortauthors{Popa L.A. et al.}

\begin{document}


\title{Cosmological parameter determination from Planck \\
and SDSS data in $\Lambda$CHDM cosmologies}


\author{L.A. Popa\altaffilmark{1}}
\affil{Institute of Space Sciences, Bucharest-Magurele, R-76900, Romania}

\author{C. Burigana and N. Mandolesi}
\affil{Istituto TeSRE, Consiglio Nazionale delle Ricerche, Via Gobetti 101,
I-40129 Bologna, Italy}


\altaffiltext{1}{further address: 
Istituto TeSRE, Consiglio Nazionale delle Ricerche, Via Gobetti 101,
I-40129 Bologna, Italy}


\begin{abstract}
We study the complementarity between the
cosmological information obtainable with the {\sc Planck}
surveyour and the large scale structure (LSS) redshift surveys in
$\Lambda$CHDM cosmologies.\\
We compute the initial full phase-space neutrino distribution
function  for $\Lambda$CHDM models by using numerical simulations.
As initial condition we adopt the HDM density fluctuation
power spectrum normalized on the basis of the analysis of
the local cluster X-ray
temperature function
and derive the initial neutrino phase-space distribution
at each spatial wave number $k$ by using the
Zel'dovich approximation.
These initial neutrino phase-space distributions
are implemented in the CMBFAST code
for the integration of the coupled linearized Einstein, Boltzmann and fluid
equations in $k$-space.
We find that the relative bias between the CMB temperature fluctuations
and the underlying matter density fluctuation power spectrum
in COBE/DMR normalization
is given by the CDM component normalized accordingly to the abundance of rich
clusters at the present time.\\
We use the Fisher information matrix approximation to
constrain  a multi-dimensional parametrization
of the $\Lambda$CHDM model, by jointly considering CMB and large scale
structure data according to the {\sc Planck} and the SDSS
experimental specifications and by taking into account redshift distortions
and nonlinear effects on the matter power spectrum.
We found that, although the CMB anisotropy and polarization
measurements tend to dominate the constraints on most of the cosmological
parameters, the additional small scale
LSS data help to break the parameter degeneracies.\\
This work has been done
in the framework of the {\sc Planck} LFI activities.
\end{abstract}




\section{Introduction}

        The simplest hypothesis for the origin of the large-scale
structure of the present universe is that it is the result of the
gravitational instability of small initial density perturbations.
In this framework,
significant improvements on
the study of the formation and evolution of the
cosmological large-scale structure have been achieved
in the recent years
\footnote{http://star-www.dur.ac.uk/virgo/virgo.html}.
In the same time, new
analysis techniques have been developed in order to extract more
sensitively the long-wavelength portion of the power spectrum
(see, e.g., Hamilton \etal 1991, Jain, Mo \& White 1995,
Peacock \& Dodds 1996, Ma 1998).
The new generation of high precision large scale structure (LSS)
redshift surveys such as the
Sloan Digital Sky survey\footnote{http://www.sdss.org} (SDSS)
and 2dF survey\footnote{http://mso.anu.edu.au/~colles/2dFGRS}
are  able to measure the total power spectrum of the matter
density fluctuations with high accuracy
for comoving wavelengts $\lambda \gsim 20$~h$^{-1}$Mpc (h=H$_0$/100 Km
s$^{-1}$Mpc$^{-1}$ is dimensionless Hubble constant), probing the strong
clustering regime effects on the power spectrum.

We are also increasingly able to probe the primordial
fluctuations through Cosmic Microwave Background (CMB) anisotropy experiments
(see, e.g., Tegmark \& Zaldarriaga 2000 for a recent compilation).
The {\sc Planck} surveyor
\footnote{http://astro.estec.esa.nl/SA-general/Projects/Planck/}
of ESA will observe the microwave sky anisotropies at frequencies between
30 and 900~GHz with FHWM angular resolutions between 33$'$ and $5'$ and
sensitivities per FWHM side squared pixel at $\sim 10 \mu$K level
at least for the channels at frequency less than $\sim 300$~GHz.

The efforts to measure the CMB anisotropy
at sub-degree angular scales and galaxy density fluctuations
in our local universe
can provide independent probes for the structure of
the universe on similar comoving scales at different cosmological
epochs [the sub-degree angular scales on the last
scattering surface of the CMB correspond to comoving scales of
$\sim 50 - 100$~h$^{-1}$~Mpc in the present local universe: an angle
$\theta$ degrees subtends a comoving distance of $105
\theta(\Omega_0 h)^{-1}$Mpc].
The comparison between these two type of measurements
is an unique way to test the
hypothesis of fluctuation growth by gravitational instability;
when the time dependence of the galaxy linear bias factor
(defined as $b_I=1/\sigma_8$, where
$\sigma_8$  represents the {\it rms} mass fluctuations in a sphere of radius
R=8~h$^{-1}$~Mpc) is understood
the combined data can be also used to constrain the parameters of
competing cosmological models
(see, e.g., Tegmark 1998, Eisenstein, Hu \& Tegmark 1998,
Eisenstein, Hu \& Tegmark 1999b, Tegmark \& Hamilton 2000).\\
The combined analysis of the CMB data and the local density field
fluctuations, as suggested by Juszkiewicz, G\'orski \& Silk (1987),
was used to obtain a corresponding CMB map
as viewed by a distant observer (Bertschinger, G\'orsky \& Dekel 1990),
starting from the density field as
derived by the POTENT procedure (Bertschinger \etal 1990, Dekel 1994).
Also, by using the likelihood analysis of the MARK III
peculiar velocity data, the density  power spectrum
for a range of parameters within the framework of CDM models
normalized to COBE/DMR  was translated
into a range of angular power spectra of the CMB anisotropy
and compared with the CMB observations  (Zaroubi et al. 1996).
The density field reconstructed via Wiener Filter method was then
translated into a map of $\Delta T/T$ as viewed by a distant observer
on his last-scattering surface.

It is difficult in the actual experimental context to explain
the observations of the cosmological large scale structure
as well as the CMB anisotropy  in
the frame of the standard Cold Dark Matter (sCDM) model normalized
to COBE/DMR data (Smoot \etal.~1992, Wright et al.~1994,
G\'orski et al.~1994, Bennett et al.~1996).
At small scales, both the  amplitude and the shape of the CMB
anisotropy power spectrum predicted by this model are inconsistent with the
observations of the Large Scale Structure (LSS) of the universe as derived
by galaxy surveys (e.g., Scott \& White 1994; White et al.~1995;
Primack \etal 1995). The small scale power excess with respect to the
large scale power can be reduced by the addition of a Hot Dark Matter (HDM)
component to the total mass density of the universe
in form of massive neutrinos  (see, e.g., Primack et al. 1995),
also motivated by the recent experimental results of the atmospheric neutrino
oscillation experiments (see, e.g., Fukuda et al. 1998, Ambrosio et al. 1998).

On the other hand, evidences have been accumulated
that we live in a low matter density universe
(see, e.g., Fukugita, Liu \& Sugiyama 1999 and the references therein).
Indications like the Hubble diagram from Type 1a
supernovae (Riess et al. 1998, Perlmutter et al. 1998) and
the acoustic peak distribution in the CMB anisotropy power spectrum
(Hancock et al. 1998, Efstathiou et al. 1999) point to a
universe dominated by vacuum energy, characterized by a cosmological constant
$\Lambda$ that keeps the universe close to be flat.
The combined analysis of the CMB anisotropy experiments
and Type 1a supernovae observations (Efstathiou et al. 1999)
indicates $\Omega_m=0.25^{+0.18}_{-0.12}$ and
$\Omega_{\Lambda}=0.63^{+0.17}_{-0.23}$ ($95$\% confidence errors)
respectively for the total matter and the vacuum energy density
normalized to the critical density,
inferring a Hubble constant value  H$_0$=65~Km~s$^{-1}$~Mpc$^{-1}$.

Current cosmological constraints on the cosmological parameters
 obtained by using the most recent CMB
anisotropy data (Lange et al. 2000, Balbi et al. 2000)
combined with Type 1a supernovae data implies a best
fit model close to a flat $\Lambda$CHDM model having
$\Omega_m \approx$ 0.33, $\Omega_{\Lambda}$=0.67 and
a neutrino density parameter
$\Omega_{\nu} \approx$ 0.1,
when the priors H$_0$=65 Km s$^{-1}$ Mpc$^{-1}$ and $\Omega_{b}$h$^2$=0.02
are assumed (Tegmark \& Zaldarriaga 2000).

The introduction of HDM component in the form of neutrinos with
the mass in the eV range supress the growth of fluctuations on all scales
below the neutrino free-streaming scale
${\rm k}^2_{{\rm fs}} =4 \pi G \rho a^2/<v>^2$
(Bond \& Sazlay 1980, Hu \& Eisenstain 1998, Ma 1999),
where $G$ is the gravitational constant, $a$ is the cosmological scale factor
($a=a_0=1$ today), $\rho$ is the density and $<v>$
is the averaged neutrino speed.
The magnitude of the power suppression is given by
$\Delta P/P \approx -8 \Omega_{\nu}/ \Omega_m
\approx -0.087 (m_{\nu}/{\rm eV}) (N_{\nu}/\Omega_m {\rm h}^2)$,
where $N_{\nu}$ is the number of massive neutrino flavours and $m_{\nu}$
is the neutrino mass.
The time dependence of the free-streaming length
implies that neutrinos cluster gravitationally on smaller
length scales at latter times (see Ma \& Bertschinger 1995, Ma 1996, Ma 1999
for discussions on the time dependence of the free-streaming scale).
In the cosmological
models involving a HDM contribution to the total energy density,
the existence of neutrino free-streaming length scale implies
that the growth of the density perturbations
depends both on  time and spatial wave number $k$.

In the linear perturbation theory, the CMB anisotropy
and  matter transfer function
are computed by the integration of coupled and linearized Einstein,
Boltzmann and fluid equations (Ma \& Bertschinger 1995) that describe
the time evolution of the metric perturbations in the perturbed density
field and the time evolution of the density fields in the perturbed spacetime
for all the relevant particle species (e.g., photons, baryons,
cold dark matter and massive neutrinos). The unperturbed energy density and
pressure of massive neutrinos as well as their
perturbed energy density and pressure, energy flux and
shear stress in the $k$-space
(see Ma \& Bertschinger 1995 for their definitions)
are  specified once the full neutrino phase-space distribution
function is known.
The full massive neutrino phase-space distribution depends on the time,
on the neutrino positions and momenta and can be represented in the form
(Ma \& Bertschinger 1995, Ma 1999):
$$f({\bf x},{\bf q},a)=f_0(\epsilon) (1+\Psi({\bf x},{\bf q},a)), \,\,\,\,\,\,\,\,\,\,
f_0(\epsilon)={g_s \over h_p^3}{1 \over {\rm e}^{\epsilon/k_BT_{\nu}}+1} \, ,$$
%
\noindent
where ${\vec q}=a{\vec p}$ is the neutrino comoving momentum,
$\vec{p}$ is the neutrino conjugate momentum,
$\epsilon=(q^2+a^2m_{\nu}^2)^{1/2}$ is the neutrino comoving energy,
$T_{\nu}=a^{-1}T_{\nu0}$ is the neutrino temperature,
$T_{\nu0}=(4/11)^{1/3} T_{\gamma0}=1.927(T_{\gamma0}/2.7{\rm K})$~K
is the present neutrino temperature, $T_{\gamma0}$ being the present
radiation temperature, $g_s$ is the number of spin for degree of freedom,
and $h_P$ and $k_B$ are the Planck and Boltzmann constants.
In the above equation $f_0(\epsilon)$ is a pure
Fermi-Dirac distribution that depends only on the neutrino comoving
energy and $\Psi$ represents a perturbation term from this distribution
depending on time and neutrino comoving momentum and position.
In the actual version of the CMBFAST code (Seljak \& Zaldarriaga 1996)
the neutrino phase-space distribution,
approximated by a pure Fermi-Dirac distribution,
does not take into account
the neutrino position dependence as indicated by the above equation.
The inclusion of the full neutrino phase-space distribution can lead to the
perturbations of the quantities related to the computation of the CMB
angular power spectrum and matter transfer function.\\
The full neutrino phase-space distribution function for Cold + Hot Dark Matter
(CHDM) models was computed before
(Ma \& Bertschinger 1994) by integrating the neutrino geodesic equations
in the perturbed background spacetime.
It was found a positive correlation between the neutrino {\it rms}
velocities and the neutrino density fluctuations at a redshift $z\sim 15$,
revealing  the contribution of the perturbations to the Fermi-Dirac
distribution.

In this paper we compute the initial full phase-space neutrino distribution
function  for CHDM and $\Lambda$CHDM models by using numerical simulations.
We start from the HDM density fluctuation
power spectrum with the normalization indicated by the analysis of
the local cluster X-ray
temperature function (Eke, Cole \& Frenk 1996)
and derive the initial neutrino phase-space distribution
at each spatial wave number $k$ by using the
Zel'dovich approximation (Zel'dovich 1970).
The neutrino phase-space distributions obtained in this way
are then implemented in the CMBFAST code
as initial neutrino momentum distributions for the
integration of the coupled linearized Einstein, Boltzmann and fluid
equations in the $k$-space and the computation of CMB power spectra and matter
transfer functions.\\
The cosmological parameters of the CHDM and $\Lambda$CHDM cosmological
models considered in this paper and the initial conditions
are presented in Section~2. The numerical
simulation approach is described in Section~3. The CMB and
matter power spectra obtained from numerical simulations
are discussed and compared with those
obtained in the case a pure Fermi-Dirac distribution in Section 4.
In Section~5 we study the implications of the nonlinear bias factor
and nonlinear effects of the matter density fluctuation power
spectra to constrain the main cosmological parameters,
when combined CMB and LSS data are take into account.
Finally, we discuss the results and draw out our conclusions
in Section~6.

\section{The initial conditions}

\subsection{Cosmological models}

We present the computation of the
full phase-space neutrino distribution function
for two distinct CHDM and $\Lambda$CHDM
models specified by the set of cosmological  parameters
given in Table~1, where we report also the parameters
of the corresponding CDM (properly sCDM and $\Lambda$CDM)
models [i.e. the sCDM ($\Lambda$CDM) model with the same parameter
of the CHDM ($\Lambda$CHDM) model but with
$\Omega_{\nu}^{{\rm sCDM}/\Lambda{\rm CDM}}=0$ and cold dark matter density parameter
$\Omega_c^{{\rm sCDM}/\Lambda{\rm CDM}}=\Omega_c^{{\rm CHDM}/\Lambda{\rm CHDM}}
+\Omega_{\nu}^{{\rm CHDM}/\Lambda{\rm CHDM}}$].
For the CHDM and the $\Lambda$CHDM models we consider
the contribution of one massless and two massive neutrino
species  while for the corresponding CDM models
we consider the contribution of
three massless neutrino species.
We assume adiabatic perturbations
and 
the presence of the scalar modes
with the spectral index $n_s=1$, as predicted by the standard
inflationary models (Guth \& Pi 1981).

It is usual to describe
the shape of the power spectrum of the density perturbations
with the shape parameter
$\Gamma=\Omega_m h \left( \Omega_r/
        \Omega_{r_0}\right)^{-1/2}$
 (Bardeen et al. 1986),
where $\Omega_r$ is the energy density parameter for all the
relativistic particles and $\Omega_{r0}$ corresponds to the
standard model with photons and three massless
neutrino species [$\Omega_{r0}=1.6813\Omega_{\gamma}$,
$\Omega_{\gamma}=2.3812 \times 10^{-5}{\rm h}^{-2}\Theta_{2.7}^4$,
$\Theta_{2.7}=T_{\gamma0}/2.7~{\rm K} \simeq 1.01$].
Observations require 0.22 $< \Gamma <$ 0.29
(Peacock \& Dodds 1994).
The treatment of the cosmological models with  HDM component
requires the introduction of a second shape parameter,
$\Gamma_{\nu}=a^{1/2} \Omega_{\nu} h^2$,
that characterizes
the effect of the neutrino free-streaming on the density fluctuation
power spectrum (Ma 1996).
Observations at the present time ($z \simeq 0$) require
$\Gamma_{\nu} <$ 0.021 (Gawiser 2000).
The values the shape parameters
$\Gamma$ and $\Gamma_{\nu}$ for our cosmological
models are also reported in  Table~1.
                         
The choice of the cosmological models presented in Table 1 is motivated
by observational evidencies. The CHDM model with two 2.4 eV
neutrinos ($m_{\nu}=92h^2\Omega_{\nu}$ eV)
was found  in remarkably agreement  with all the
available LSS observations if h$\simeq 0.5$ (Primack et al. 1995).
The $\Lambda$CHDM model
was found to be the best fit model of the current CMB
experimental data combined with the Type 1a
supernovae data when priors on H$_0$ and $\Omega_b$ are assumed
(Tegmark \& Zaldarriaga 2000).

\begin{table}[]
\caption[]{Parameters of the considered cosmological models and other
relevant quantities ($\Omega_m=\Omega_b+\Omega_c+\Omega_{\nu}$).}
\label{tableDummy}
\begin{flushleft}
\begin{tabular}{cccccccccccccc}
\hline\\
Model& $\Omega_b$&$\Omega_c$&$\Omega_{\nu}$&$\Omega_{\Lambda}$&
h& $\Gamma$& $\Gamma_{\nu}$&$\sigma_8^{\rm HDM}$&$\sigma_8$&$\sigma_8^c$&
$k_{fs}$ & $k_{nl}$  \\
 & & & & & & & & & & & (Mpc$^{-1}$) & (Mpc$^{-1}$) \\
\hline\\
CHDM &0.05& 0.75& 0.2& 0& 0.5&0.58&0.05&0.68&0.36$\pm$0.02&0.81&0.06&0.46\\
$\Lambda$CHDM&0.03&0.2&0.1&0.67&0.65&0.25&0.04&0.33&0.32$\pm$0.02&0.31&0.03
&0.35\\
sCDM&0.05&0.95&0&0&0.5&0.5&-&-&0.52$\pm$0.03&1.17&-&0.46\\
$\Lambda$CDM&0.03&0.3&0.&0.67&0.65&0.21&-&-&0.88$\pm$0.06&0.92&-&0.35\\
\hline
\end{tabular}
\end{flushleft}
\end{table}[]

\subsection{Matter density fluctuation power spectrum and its normalization}

\begin{figure}
\plotone{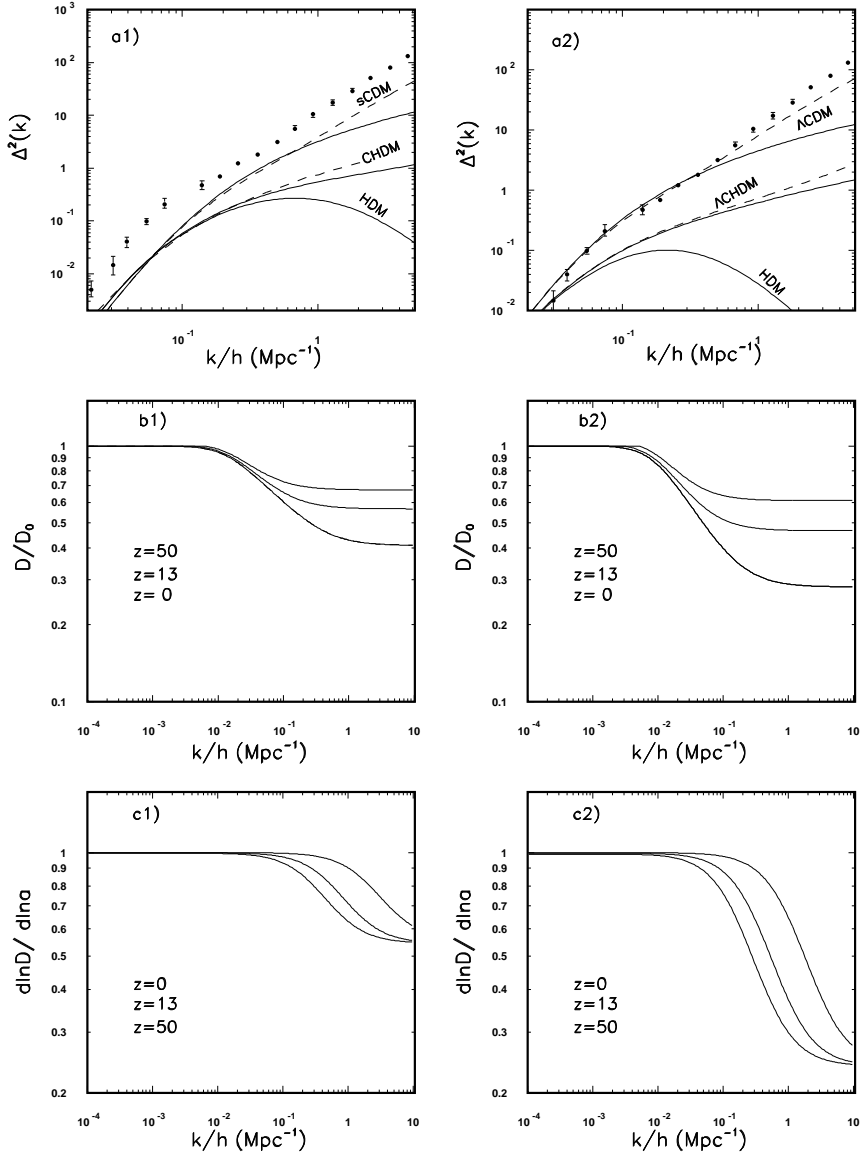}
\caption{Panels a1) and a2):
the weighted matter density
fluctuation power spectra and the HDM component density fluctuation
power spectra. The power spectra in the nonlinear regime are indicated with
dashed lines. All the power spectra are normalized at $\sigma_8$ and
are computed at the present time.
The experimental data points are derived from the APM galaxy survey.
The evolution of the weighted matter density
fluctuations and of the HDM  component density fluctuation
is shown in panels b1), c1) and b2), c2) respectively.
Panels b1) and b2): the growth functions at some redshift values
as function of the spatial wave number $k$.
Panels c1) and c2): the corresponding growth rates at some redshift values
as function of the spatial wave number $k$.}
\end{figure}

The power spectrum of the matter density fluctuations is defined as
(Bardeen et al. 1996):
\begin{equation}
{\cal P}(k,a)={\cal A} k^{n_{s}} T(k,a)^2 \left( \frac{D(k,a)}{D_0(k)}
\right)^2 \, ,
\end{equation}
where ${\cal A}$ is a normalization constant,
$k$ is the spatial wave number,
$T(k,a)$ is the matter transfer function and  $D(k,a)$
is the growth factor of the density perturbations assuming a value
$D_0(k)=D(k,a=1)$ at the present time.
We will refer in the following also to the growth rate, defined as
$f(k,a)={\rm d ln D}/{\rm dln a}$.

The dimensionless power spectrum $\Delta^2(k,a)$,
defined as the power
per logarithmic interval in spatial frequency
$k$ (the power variance),  is given by (Peebles 1980):
\begin{equation}
\Delta^2(k,a)=\frac{1}{2 \pi^2} k^{n_s+3}{\cal P}(a,k).
\end{equation}
The total matter density fluctuations contributed by the
different matter components
(baryons, cold dark matter particles, neutrinos)
can be obtained by replacing $T(k,a)$
in the equation (1) by the
the weighted transfer function given by (Ma 1996):
\begin{equation}
T(k,a)=\frac{\Omega_b}{\Omega_m} T_b(k,a)+
         \frac{\Omega_{c}}{\Omega_m} T_{CDM}(k,a) +
        \frac{\Omega_{\nu}}{\Omega_m} T_{\nu}(k,a),
\end{equation}
where $T_b(k,a)$,  $T_{CDM}(k,a)$ and  $T_{\nu}(k,a)$ are the
transfer functions of baryons, cold dark matter and neutrinos respectively.
The growth factor
determines the normalization of the amplitude of matter density fluctuations
relative to the CMB one (Eisenstein, Hu \& Tegmark 1008).
For the sCDM and $\Lambda$CDM models
the growth functions depend only on
time (see Lahav \etal 1991, Carroll, Press \& Turner 1992).
The presence of  neutrino free-streaming
alters the growth rate and then for CHDM and $\Lambda$CHDM models
the growth factors and the growth rates
depend on the time and the spatial wave number $k$
(Hu \& Eisenstein 1998).

It is usual to define the normalization of the density
fluctuation power spectrum as $\sigma_8(a)$, the linear
{\it rms} mass fluctuations in a sphere of radius R=8$h^{-1}$Mpc,
since the observed {\it rms} galaxy
counts on this scale is about unit.
$\sigma_8(a)$ is related to the power spectrum $\Delta^2(k,a)$
through:
\begin{equation}
\sigma_8(a)=\left[ \int_0^{\infty}\frac{dk}{k}
\Delta_i^2(k,a)W^2({\rm x}) \right]^{1/2},
\end{equation}
where ${\rm x}=k$R and  $W({\rm x})$ is the window function.
We take a top-hat window function
of radius R=8h$^{-1}$Mpc in the real space:
\begin{equation}
W({\rm x})= 3(\sin {\rm x} -{\rm x}\cos{\rm x}) / {\rm x}^3.
\end{equation}
It is also possible to set the amplitude of the initial
fluctuation spectrum
to the measured CMB temperature fluctuations on large scales provided
by the COBE/DMR experiment (Bunn \& White 1997).
However, the COBE
normalization is sensitive to the power spectrum at
$k \simeq 10^{-3} h$~Mpc$^{-1}$, that is not relevant for the galaxy
clustering. Alternatively to the COBE normalization, the abundance of rich
clusters can be used to fix
the amplitude of the initial
mass fluctuations close to the quasilinear scale
(see also Jenkins \etal 1998 and Peacock 2000
for a discussion of the advantages of this kind of normalization).

We adopt the values of
$\sigma_8$ obtained by Eke, Cole \& Frenk 1996 for CDM models
as derived from the analysis of the local X-ray temperature function
(for $\sigma_8$ values obtained from slightly
different analyses and  consistent with these values
see also  White, Efstathiou \& Frenk 1993,
Viana \& Liddle 1996, Pen 1997):
\begin{eqnarray}
\sigma_8^{{\rm CDM}}&=&(0.52 \pm 0.04)\Omega_m^{-0.52+0.13 \Omega_m}
\hspace{0.2cm} {\rm (flat \; models)} \\
\sigma_8^{{\rm CDM} }&=&(0.52 \pm 0.04)\Omega_m^{-0.46+0.1 \Omega_m}
\hspace{0.2cm} {\rm (open \; models)} \nonumber
\end{eqnarray}
For the cosmological models involving a HDM component,
the value of $\sigma_8$ is given by (Ma 1996):
\begin{equation}
\sigma_8=\sigma_8^{{\rm CDM}} \times
          \sigma_8^{\rm HDM},
\end{equation}
where  $\sigma_8^{\rm HDM}$ and $\sigma_8^{ {\rm CDM} }$
are the normalizations for
the HDM model and the corresponding CDM model respectively.
For the purpose of this paper we take
$\sigma_8^{{\rm CDM}}$ as given by the equations (6) and
$\sigma_8^{{\rm HDM}}$ as obtained in the COBE normalization,
as computed by the CMBFAST code.
The values  of $\sigma_8$ and
$\sigma_8^{{\rm HDM}}$ for the cosmological models considered
here are also reported in Table~1.
For comparison, we  present also  the values
of $\sigma_8^c$ derived when
$\sigma_8^{\rm CDM}$ is obtained in the COBE normalization.
One can see that while for $\Lambda$CHDM models
the $\sigma_8$ values derived on the basis of these two normalizations
are almost identical, they differ by a factor of about two
in the case of CHDM models.\\
We also indicate in Table~1 the values of the free-streaming wave number
$k_{fs}$ and of the wave number $k_{nl}$ at which the
nonlinear effects become important (Peacock \& Dodds 1996):
\begin{eqnarray}
k_{{\rm fs}} &\approx& 0.026 \left(\frac{m_{\nu}}{1{\rm eV}}\right)^{1/2}
\Omega^{1/2}_m
{\rm hMpc}^{-1}, \\
k_{nl}&=&\left[\frac{((n_{{\rm eff}}+1)/2)!}{2}\right]^{1/(n_{{\rm eff}}+3)}
\frac{\sqrt{10}}{R}{\rm hMpc}^{-1},
\hspace{0.3cm} n_{{\rm eff}}(k_L)=\frac{{\rm d ln }P(k)}{{\rm d ln} k}
(k=k_L/2) \nonumber,
\end{eqnarray}
where $k_L$ is the linear wave number and
$n_{{\rm eff}}(k_L)$ is the effective spectral index of the power spectrum.\\
Figure~1 presents the  weighted
matter density fluctuation power spectra and the HDM component
density fluctuation power spectra for the considered cosmological models,
with the transfer functions given by  the CMBFAST code.
All the  matter power spectra are normalized
at $\sigma_8$ as given by the  equations (6)-(7)
and are computed at the present time.

The nonlinear  matter density fluctuation power spectra
(dashed lines in Figure~1) are computed with the formalism of
Peacock \& Dodds 1996 by using the appropriate linear growth rate
functions given by
the ratio between the power spectrum
of each cosmological model
and the power spectrum of the corresponding CDM model (Ma 1996)
calculated with the linear growth factor of the CDM model given by
(Peebles 1980):
\begin{eqnarray}
g(\Omega_m,\Omega_{\Lambda})=\frac{D}{a}=
\frac{5\Omega_m}{2}\int_0^1 \frac{{\rm d}a}{a^3H(a)^3},
\end{eqnarray}
where $\Omega_K=1-\Omega_m-\Omega_{\Lambda}$ is the curvature density parameter
and
$H(a)=~(\Omega_m a^{-3}+\Omega_K a^{-2}+\Omega_{\Lambda})^{1/2}$
is the Hubble expansion rate  normalized to unit at the present time.
We test the validity of Peacock \& Dodds formula
for $k/h \sim 5$~Mpc$^{-1}$, finding a very good agreement with the
analytical approximation to the nonlinear power spectra for CHDM models
obtained by Ma (1998).\\
We present also in Figure~1  the corresponding
growth functions and growth rates for the considered cosmological
models computed at some representative epochs.

\section{N-body simulations}

We obtain the initial neutrino phase-space distributions
at each spatial wave number $k$
through numerical simulations based on
the standard particle-mesh
(PM) method (Efstathiou \& Eastwood 1981, Hockney \& Eastwood 1981)
usually used to set the initial conditions
for the nonlinear evolution of the large-scale structure
(see Jenkins et al. 1998 and the references therein).

The initial neutrino positions and velocities
are generated from the HDM matter density fluctuation power spectrum
with the normalization given by the equation (6)-(7) by using
Zel'dovich approximation (Zel'dovich 1970).
According to the Zel'dovich  approximation, the perturbed comoving
position of a particle ${\vec r}( {\vec r_0}, a)$ and its peculiar
velocity ${\vec v}( {\vec r_0},a)$ are related to the
fluctuations of the density field
$\delta \rho_{HDM}({\vec r_0},a,k)$ through:
\begin{eqnarray}
{\vec r}( {\vec r_0},k,a) &=& {\vec r_0}+D(k,a)
{\vec d}({\vec r_0}) \, , \;\;\;  
{\vec v}( {\vec r_0},k,a) = {\dot D}(k,a){\vec d}({\vec r_0}) \, , \\
{\vec \nabla}{\vec d}({\vec r_0})
&=& D^{-1}(k,a) \delta \rho_{HDM}({\vec r_0},k,a) \nonumber \, , 
\end{eqnarray}
where ${\vec r_0}$ is the coordinate corresponding to the
unperturbed comoving position
and ${\vec d}({\vec r_0})$ is the displacement field.
In the simulations we use the set of equations (10)
to compute the perturbed neutrino comoving positions,
the neutrino peculiar velocities and the
displacement of the density fields
for each wave number $k$ at the present time.
We assign to each neutrino a momentum according to the growth
function, when the power of each mode is randomly selected
from a Gaussian distribution with the mean accordingly
to the  power spectrum of the HDM component
(Hoffman \& Ribak 1991, Ganon \& Hoffman 1993, Bertschinger 1995), and
add a thermal momentum randomly drawn from a Fermi-Dirac
distribution:
\begin{eqnarray}
 f(q_0) \sim \frac{ q_0^2}{e^{q_0}+1},
\hspace{0.3cm} q_0={\cal M}_{\nu}v/c \hspace{0.3cm}
{\rm and } \hspace{0.3cm}
{\cal M}_{\nu}=\frac{m_{\nu}c^2}
{k_B T_{\nu_0}},
\end{eqnarray}
where ${\vec v}$ is the neutrino velocity and $c$ is the speed of light.
We performed simulations with $10 \times 32^3$, $10 \times 64^3$
and $ 10 \times 128^3$ particles.
The neutrinos with identical masses are randomly placed
on $32^3$, $64^3$ and $128^3$ grids, 10 per grid point,
with  comoving spacing $r_0$ in the range (0.5~--~5)~h$^{-1}$Mpc.
We verify the convergence of the results with the variation  of the
number of particles and the comoving spacing.
The results presented here are obtained from ten simulations
with $10 \times 128^3$ neutrinos
and a comoving spacing of $r_0$=0.5Mpc.
This large number of particles ensures
enough statistics for the computation of the phase-space
distributions, while this comoving spacing was found to
give the minimum variance
of the likelihood distribution functions presented in Figure 4.
In the computation of the set of equation (10) we consider
only the growing modes,
the nonlinear power spectrum up to $k_{max}=6.28$~h~Mpc$^{-1}$
and neglect the contribution of the redshift distortions on the power
spectrum.

The neutrino momentum field obtained at each wave number $k$
was sampled in fixed equispaced
points (we use here $N_{q_{max}}$=50) and normalized to
the neutrino total number
($N_{part}=10 \times 128^3$ in the current simulation).

Figure 2 shows (left panels)
the contour plots of the constant particle probabilities
in the $\delta q$-$\delta \rho_{HDM}$ plane, where $\delta q=q_i(k)-<q(k)>$,
$q_i(k)$ being the peculiar neutrino momentum and $<q(k)>$
the bulk neutrino momentum of the lattice:
\begin{equation}
< q(k) >=\frac{1}{N_{part}} \sum_i  q_i(k).
\end{equation}
We also show (right panels)
the momentum distribution
functions obtained from numerical simulations
(continuous line)
compared with the
thermal momentum distribution function
(dashed line). We find a
displacement of the bulk neutrino momentum
towards higher values for the same wave number when
$\sigma_8$ increases. Also, for the same value of $\sigma_8$,
the bulk neutrino  momentum is displaced towards
smaller values with the increasing of the wave number, reflecting
the $k$-dependence of the growth function. The dominant effect is given
by the variation of the $\sigma_8$ value.

\begin{figure}
\plotone{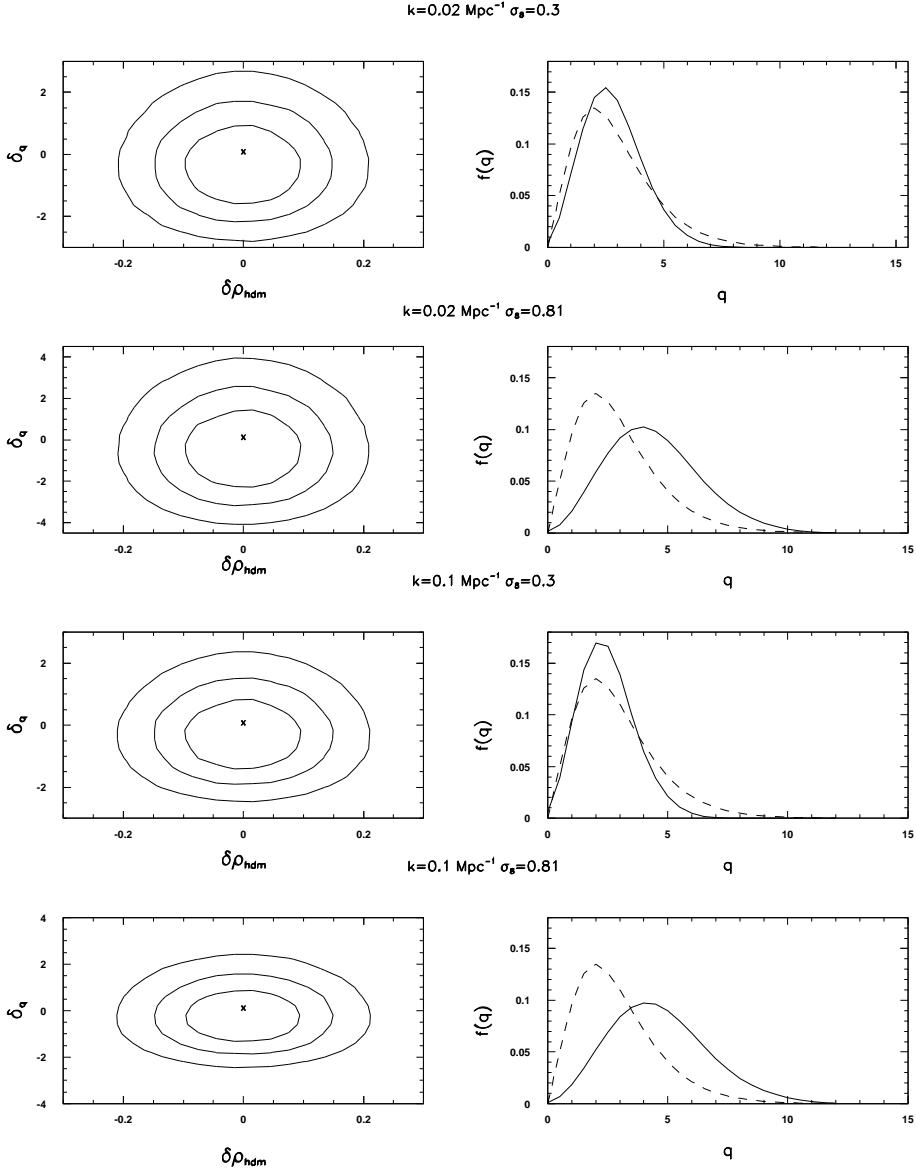}
\caption{Left panels: the contour plots of constant particle probabilities
in the
$\delta q - \delta \rho_{HDM}$ plane (see also the text).
From exterior to interior the contours correspond
to: 0.75, 0.5 and 0.25 probability.
Right panels: the momentum distribution functions obtained from
numerical simulations (continuous line) and
the thermal momentum distribution function (dashed line).
These plots refer to the case of the CHDM cosmological model.}
\end{figure}

\begin{figure}
\vskip -1cm
\plotone{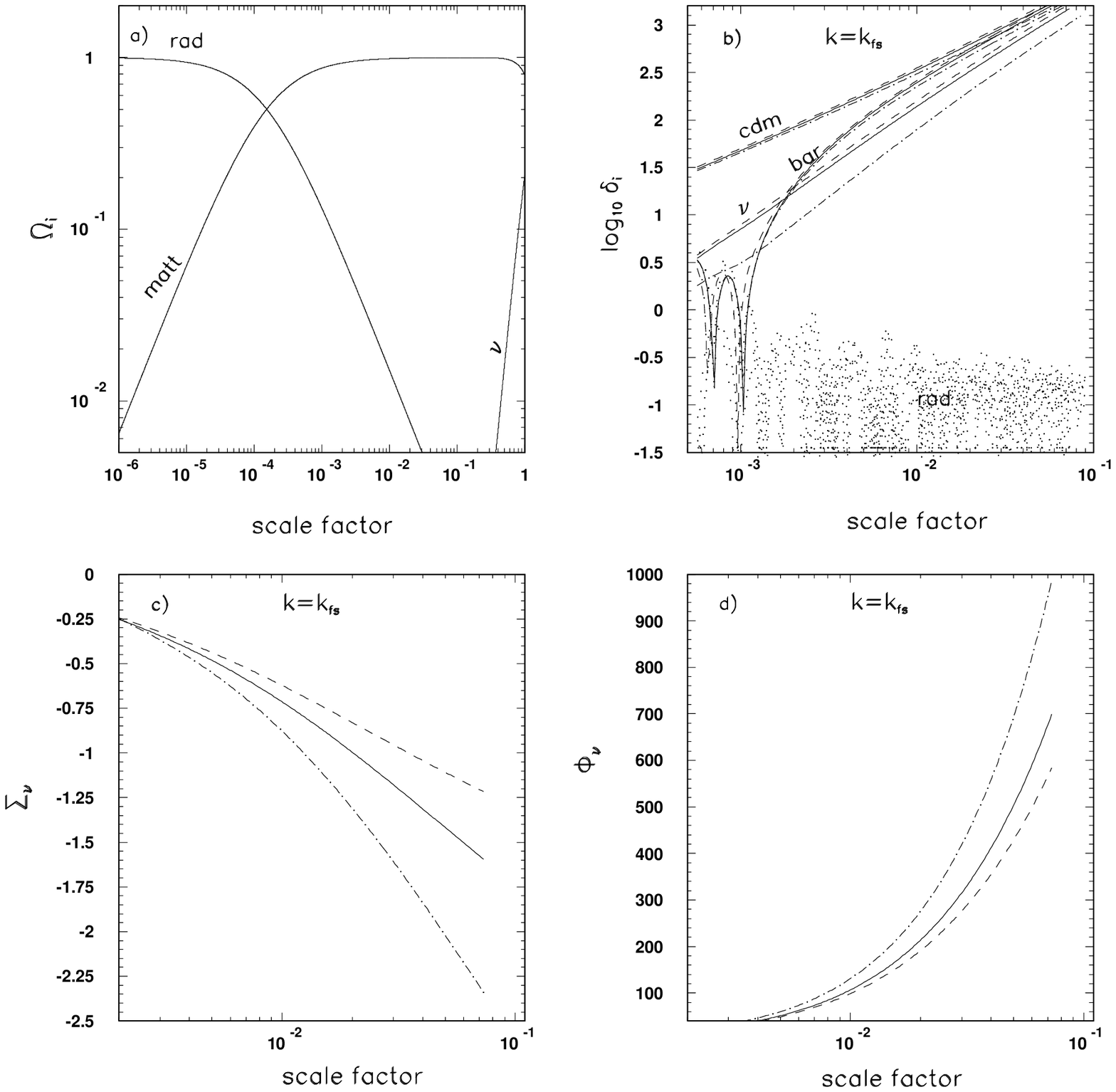}
\caption{Panel a): the time evolution of the energy density of
the different components (see also the text).
Panel b): the time evolution of energy density perturbations
for different components when the neutrino phase-space distribution
is a Fermi-Dirac distribution (continuous line) or it 
is obtained from numerical simulations with $\sigma_8=0.3$
(dashed line) or with $\sigma_8=0.81$ (dotted-dashed line).
Panel c): the time evolution of the neutrino shear stress
when the neutrino phase-space distribution
is a Fermi-Dirac distribution (continuous line) or it is
obtained from numerical simulations with $\sigma_8=0.3$
(dashed line) or with $\sigma_8=0.81$ (dotted-dashed line).
Panel d): the same as in the panel c) but
for neutrino energy flux.
The distributions presented in panels b) c) and d) are obtained at
$k=k_{fs}$.
All the plots refer to the CHDM cosmological model.}
\end{figure}

\section{The CMB power spectrum from numerical simulations}

The neutrino momentum distributions obtained from numerical simulations
at each wave number $k$
are used as initial neutrino phase-space distributions
in the CMBFAST code for the integration of the Boltzmann, Einstein
and fluid equations. As usual,
the  neutrino comoving energy is then
$E=(q^2+a^2{\cal M}^2_{\nu})^{1/2}$ (Ma \& Bertschinger 1995),
where $q$ is the neutrino comoving momentum.
Following the same procedure implemented in the CMBFAST code,
we compute in synchronous gauge the perturbations of the
energy density, pressure, energy flux and shear stress,
truncating the Boltzmann hierarchies
for massive neutrinos
at $l_{max}=50$ for every value of $q$, obtaining
a relative accuracy in the estimation
of the CMB power spectrum better than
$\simeq 10^{-3}$ (Popa et al. 2000).

Figure 3 presents the evolution with the scale factor
of the energy density parameters of different components and of
their density perturbations, and of the neutrino shear stress and energy flux
obtained for  the CHDM model for the neutrino phase-space distribution
function modelled as a Fermi-Dirac distribution (continuous lines)
or obtained from the numerical
simulations when the normalization of the matter power spectrum is
$\sigma_8=0.3$ (dashed lines)
or $\sigma_8=0.81$ (dotted-dashed lines).
One can see that while the time evolution
of the various energy density components
obtained from numerical simulations
are not changed, their perturbations as well as the
neutrino shear stress and energy flux differ from  those
obtained in the case of a pure Fermi-Dirac distribution.
As it was mentioned before, the full neutrino phase-space distribution
depends also on the neutrino position, leading to perturbations
of the pure Fermi-Dirac distribution,
$f({\vec x},q,t)=f_0(q)[1+\Psi({\vec x},q,t]$, where
$f_0(q)$ is the pure Fermi-Dirac distribution.
This perturbations are reflected by the time evolution
of the quantities presented in  Figure~3.

Figure 4 presents the CMB anisotropy power spectra
obtained from numerical simulations
for different normalizations
of the matter density fluctuations power spectra,
for the CHDM (panel a1))
and the $\Lambda$CHDM models (panel b1)). Each power spectrum
is obtained by averaging the power spectra  from ten simulations
with $10\times 128^3$ particles
and normalized to COBE/DMR  (Bunn \& White 1997).
We also show in  Figure 4 the likelihood
dependence on $\sigma_8$ for the CHDM model (panel a2))
and $\Lambda$CHDM model (panel b2)).
For each case the target power spectrum  is the corresponding
power spectrum given by
the CMBFAST code normalized to the COBE/DMR data.
We found  at $1-\sigma$ level the following values for $\sigma_8$:
\begin{eqnarray}
\sigma_8=0.34 \pm 0.09 \hspace{0.3cm} {\rm for}
\hspace{0.1cm}{\rm CHDM} \nonumber \\
\sigma_8=0.3 \pm 0.07 \hspace{0.3cm} {\rm for}
\hspace{0.1cm} \Lambda{\rm CHDM} \, . \nonumber
\end{eqnarray}
These values of $\sigma_8$ show
that the CMB power spectrum in COBE normalization
is well  recovered when the matter density fluctuation power spectrum
is normalized to the cluster abundancy data, as indicated in the Table~1.
The results presented in Figure~4 show that the relative amplitude
between the CMB temperature fluctuations and the matter density
fluctuations normalized to COBE/DMR
is well defined by the CDM component with the normalization
indicated by the abundance of the rich clusters at the present time.
This result confirms the predictions of the structure formation theories
in which the CDM driven by the adiabatic fluctuations leads to the formation
of the CMB anisotropy and large scale structure
(see, e.g., Dodelson, Gates \& Turner 1996).
\begin{figure}
\plotone{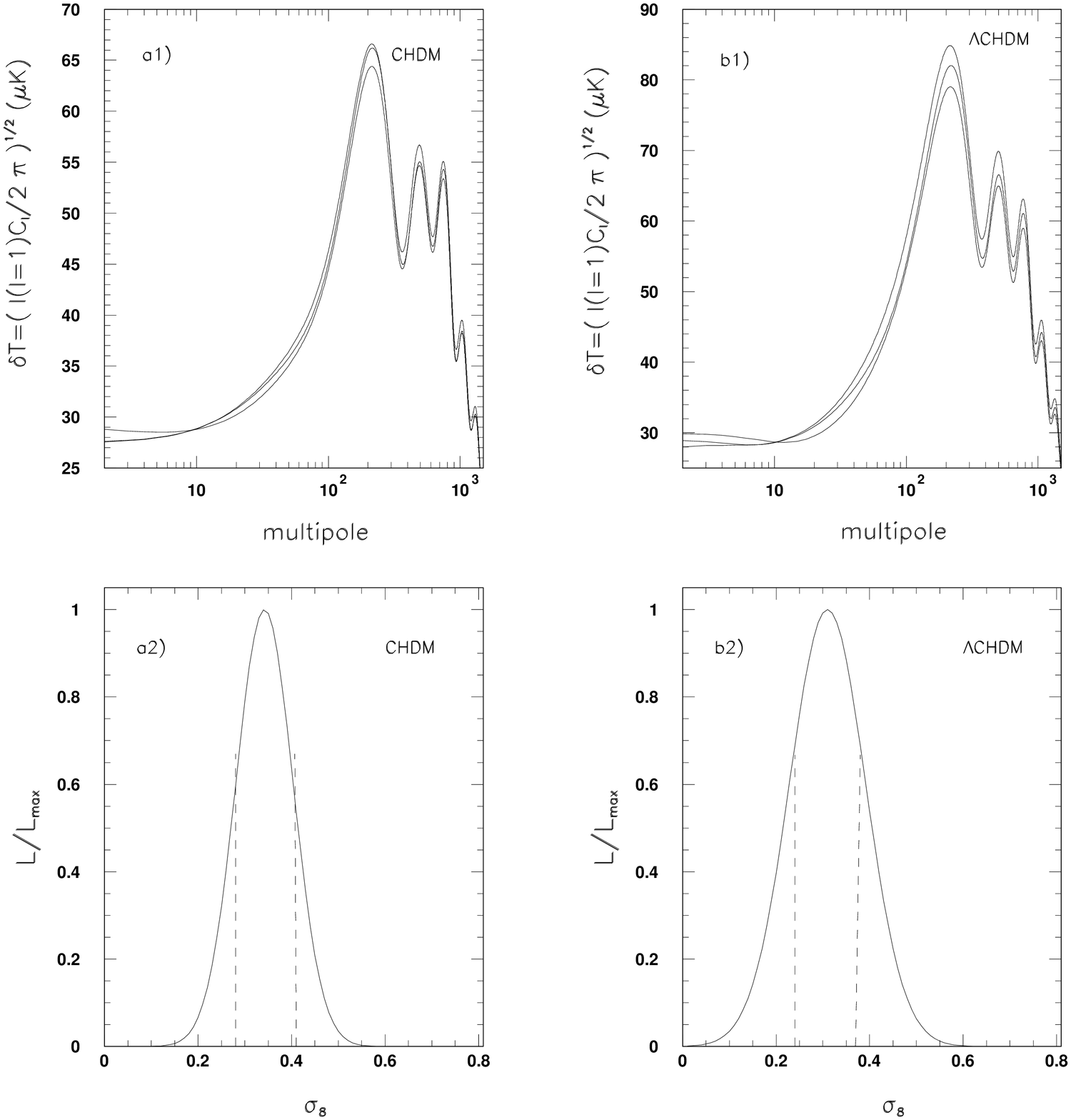}
\caption{The CMB anisotropy power spectra obtained from
numerical simulations for CHDM model (panel a1))
and $\Lambda$CHDM model (panel a2)) for
different values of $\sigma_8$.
From the top to the bottom (at the first Doppler peak):
$\sigma_8$=0.2, 0.3, 0.5.
The likelihood dependence on $\sigma_8$ obtained
in CHDM model (panel b1))
and $\Lambda$CHDM model (panel b2)).
All the reported power spectra
are obtained averaging over ten simulations and are normalized
to COBE/DMR data.}
\end{figure}
Figure 5 presents  the time evolution of
the bias factor $b_I=1/\sigma_8$ (panel a))
and of the parameter $\Omega_m^{0.6}/b_I$ (panel b)), usually measured
from the peculiar velocity data, obtained for the CHDM and the $\Lambda$CHDM
model.
For both models the normalization at the present time is given by the
equations (6)-(7).
\begin{figure}
\plotone{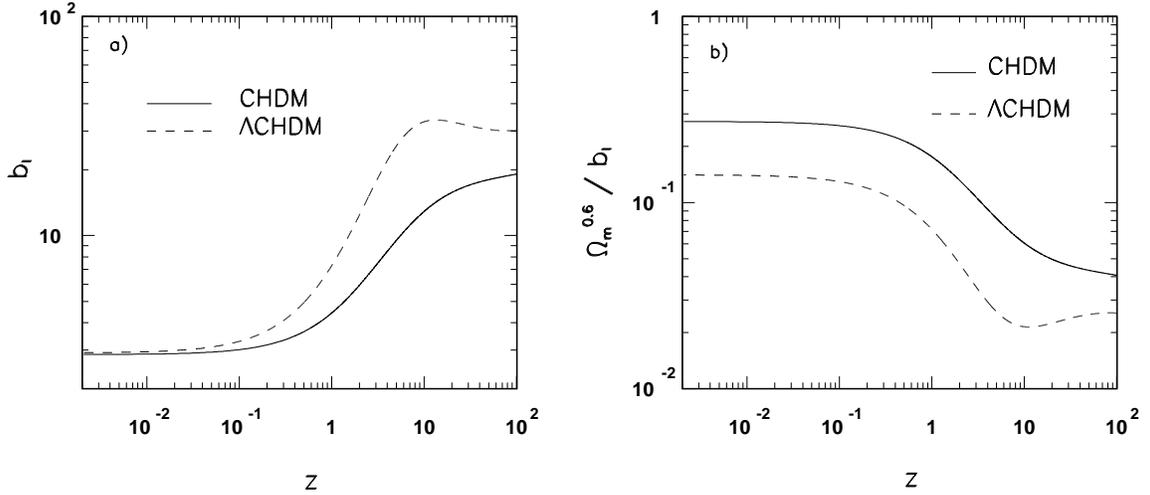}
\caption{The redshift dependence of the bias factor $b_I$ (panel a))
and of the parameter $\Omega_m^{0.6}/b_I$ (panel b)) obtained
for the CHDM and the $\Lambda$CHDM models, when the CDM component
at the present time
is normalized to the cluster abundancy data (see also the text).}
\end{figure}

\section{Cosmological parameter determination  from CMB
and LSS data in $\Lambda$CHDM cosmologies}

The results obtained in the previous section show that
the relative bias between the CMB anisotropy
power spectrum
and the matter density fluctuation power spectrum in the COBE/DMR
normalization is given by the CDM component with the normalization obtained
from the abundance of the reach clusters at the present time.
This confirms the idea to jointly use
the cosmological information contained in
the CMB power spectrum and in the matter power spectrum
to constrain degenerated sets of cosmological parameters.

There are few factors  that  alter the
relative amplitude between CMB temperature fluctuations and
the underlying mass density fluctuations:
the  growth rate of perturbations depends upon
$\Omega_m$ and $\Omega_{\Lambda}$ and implicitly
upon $\Omega_K$ (see equation (9)).
The degeneracy between $\Omega_{\Lambda}$ and $\Omega_K$ leeds to
an indeterminacy of the Hubble constant (Efstathiou \& Bond 1999):
\begin{eqnarray}
h=(\omega_m+\omega_{\Lambda}+\omega_k)^{1/2},
\end{eqnarray}
where $\omega_i=\Omega_i {\rm h}^2$ are the physical densities of the
relevant components. The CMB anisotropy measurements can constrain
$\omega_m=\omega_b+\omega_c+\omega_{\nu}$ and
$\omega_b$  from the morphology of the Doppler peaks.

On the other hand, the observed power spectrum in the redshift space
differs from the
theoretical power spectrum of mass fluctuations because of
the nonlinear evolution, reshift-space mapping and the bias.
For the purpose of this paper we consider a scale-independent bias factor
$b_I=1/\sigma_8$, and compute  the effects of the nonlinear evolution
by using the Peacock \& Dodds formula (see Section~2). Following
Feldman, Kaiser \& Peacock (1994) we correct the nonlinear power
spectrum of mass fluctuations for redshift distortions.
On large scales the redshift distortions cause a linear
increase of the power given by:
\begin{equation}
{\cal P}(k) \rightarrow {\cal P}(k) \left[\frac{\Omega^{0.6}_m}{3b_I}+
\frac{\Omega_m^{1.2}}{5b_I^2}\right].
\end{equation}
On small scales the redshift distortions cause:
\begin{equation}
{\cal P}(k) \rightarrow {\cal P}(k) \frac{\sqrt {\pi }}{2}
\frac{{\rm erfc}(k \sigma)}{k \sigma},
\end{equation}
where $\sigma$ measures the spatial {\it rms} distortion and erfc is the
complementary error function. The typical values of $\sigma$,
indicated by the {\it rms} small scale
velocity measurements (Feldman, Kaiser \& Peacock 1994), is of few
h$^{-1}$Mpc.
The indeterminacy of h
causes the movement of the matter density power spectrum in the redshift space,
changing the normalization scale, the neutrino free-streaming scale
$k_{{\rm fs}}$, and the scale $k_{{\rm nl}}$ at which the nonlinear
effects become important  (see Section~2).\\
The relative normalization between the CMB and matter power spectra
depends also on $\omega_m$.
While $\omega_m$ and $\omega_b$ can be constrained
from the CMB Doppler peak distribution, the strong dependence of
the matter power spectrum on  $\omega_{\nu}$ at intermediate and small
scales constrains $\omega_c$.\\
The spectral index of the scalar modes $n_s$, the spectral index of
the tensorial modes $n_t$,
the ratio T/S between tensorial and scalar contributions as well as
the reionization effects (an optical depth to the last scattering
$\tau~ \neq~ 0$~)
affect  also the relative bias between the CMB temperature fluctuations and
the matter density fluctuations.
Within the CMB polarization data $n_t$, $T/S$
and $\tau$ can well be constrained,
while the indeterminacy of most of the
cosmological parameters is reduced
(see, e.g., Zaldarriaga, Spergel \& Seljak  1997, Zaldarriaga \& Seljak 1997).

We compute the precision at which the fundamental cosmological
parameters could be extracted from the combined CMB anisotropy and
LSS data given by {\sc Planck} and SDSS, 
taking into account the dependence of the
CMB and matter power spectrum on all
the factors enumerated before.
The previous papers on the parameter
estimation from CMB and LSS data
(see, e.g., Scott \etal 1995, Bond, Jaffe \& Knox  1998,
Lineweaver 1998, Webster \etal 1998, Wang \etal 1998, Eisenstein, Hu \& Tegmark 1999b)
in general consider the
normalization to the COBE data,
neglecting the cosmological contribution to the estimation of the
cosmological parameters  of the nonlinear effects and  redshift distortions.

We use the Fisher  matrix approximation
to compute the errors on the estimates
of the cosmological parameters (see, e.g., Efstathiou \& Bond 1999,
Popa et al. 1999) for the experimental specifications of {\sc Planck}
and SDSS.
In this case, the Fisher matrix elements are  the sum of two terms,
accounting for the Planck measurements of the CMB anisotropy
and polarization and the galaxy
power spectrum as derived from SDSS (Tegmark 1997):
\begin{equation}
F_{ij}=F_{ij}^{\rm Planck}+F_{ij}^{\rm SDSS}.
\end{equation}
The minimum error that can be obtained
on a parameter $s_i$ when we need to determine all parameters
jointly, is given by:
\begin{equation}
\delta s_i = \sqrt{F^{-1}_{ii}},
\end{equation}
depending on the experimental specifications, the target model
and the class of considered  cosmological models.
If  both anisotropy and polarization power spectra are used,
the first Fisher information matrix term is given by (Zaldarriaga \&
Seljak 1997, Zaldarriaga 1997):
\begin{equation}
F^{\rm Planck}_{ij}=\sum_{l} \sum_{X,Y} \frac{\partial C_{Xl}}{\partial
s_{i}}
              Cov^{-1}({\hat C}_{Xl},{\hat C}_{Yl})
                            \frac{\partial C_{Yl}}{\partial s_{j}},
\end{equation}
where $X$ and $Y$ stands for $T$, $E$, $C$ and $B$ power spectra and
$Cov^{-1}$ is the inverse of the covariance matrix.
The {\sc Planck} experimental parameters and the procedure used to compute
the Fisher matrix term given by the equation (18) are described in
Popa et al. (2000); in particular, we consider here, for
simplicity, the {\sc Planck} ``cosmological channels''
only (between 70 and 217~GHz) and neglect the foreground contamination.  \\
The second term of the Fisher information matrix
in equation (16) is given by (Tegmark 1997):
\begin{equation}
F^{{\rm SDSS}}_{ij}=2 \pi \int_{k_{min}}^{k_{max}} \frac{\partial
{\rm ln}P(k)}{\partial s_i}
w(k)
\frac{\partial {\rm ln }P(k)}{\partial s_j}{\rm d ln}k,
\end{equation}
where $k_{min}$, $k_{max}$ are the minimum and maximum wave number
used to compute the matter density fluctuation power spectra
(we take a fixed $k_{min}=10^{-4}$hMpc$^{-1}$  while
the value of $k_{max}$ can vary)
and $w(k)$ is the selection function for the Bright Red Galaxy (BRG)
sample of the SDSS.  According to Tegmark (1997):
\begin{equation}
w(k)=\frac{V_{eff}(k)}{\lambda^3}.
\end{equation}
Here $V_{eff}(k)$ is the effective volume of
the BRG  sample used for measuring
the power at the wave number $k$ corresponding to the wavelength
$\lambda=2 \pi/k$:
\begin{equation}
V_{eff}=\int \left[ \frac{{\bar n}({\vec r}{\cal P}(k)}{1+
{\bar n}({\vec r}) {\cal P}(k)}\right]{\rm d}^3 r,
\end{equation}
where ${\bar n}({\vec r})$ is the selection function of the
survey that gives the expectation value of the number density of galaxies.
We consider the BRG sample of SDSS volume-limited at 1000h$^{-1}$Mpc,
containing 10$^5$ galaxies (Tegmark 1997) with a bias factor given by
the equations (6)-(7) and ${\bar n}({\vec r})$ is the
expectation value of the
Poisson distribution.

\begin{figure}
\vskip -5cm 
\plotone{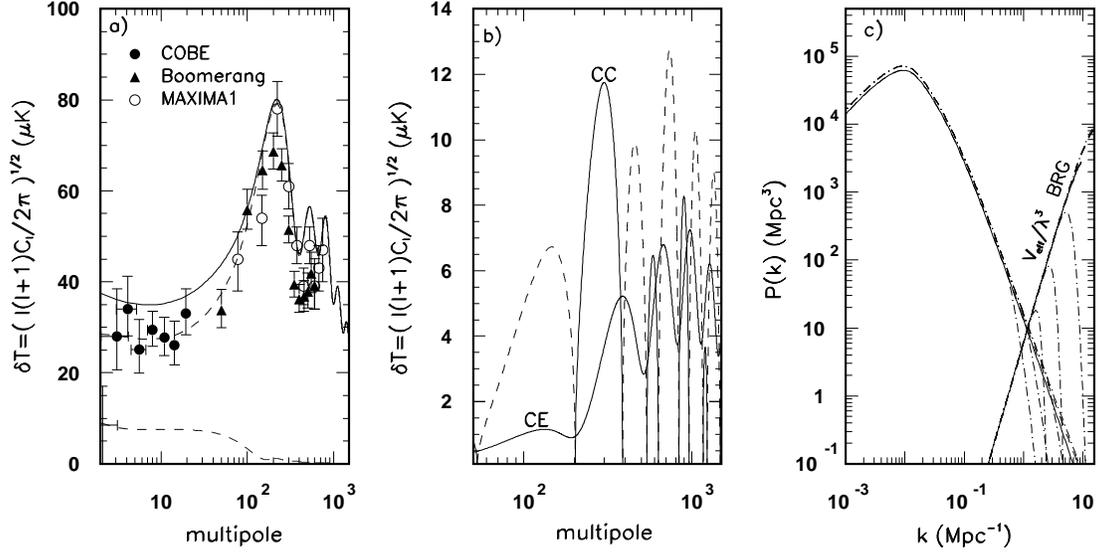}
\caption{Panel a): the CMB anisotropy power spectrum of the
fiducial model (continuous line) and its decomposition into scalar
and tensorial contributions (dashed lines),
compared with the CMB anisotropy experimental data from
COBE, Boomerang and MAXIMA1.
Panel b): the polarization power spectra of the fiducial model
for E-channel (continuous line) and cross-correlation C-channel
(dashed and dotted-dashed lines; the dotted-dashed line represents the
negative part of the power spectrum).
Panel c): the matter fluctuations
power spectrum of the fiducial model
and $V_{eff}(k)/\lambda^3$ obtained for the
fiducial model and the experimental
specifications of the BRG sample of SDSS.
The solid curves are for the linear regime,
the dashed curves are obtained for the
nonlinear regime and the dotted-dashed curves are obtained when both
nonlinear effects and redshift distortions are considered,
for few choices of $\sigma$; from top to bottom
$\sigma$=0,0.2,0.5,1 $h^{-1}$Mpc (see also the text).
The power spectra are computed at the present time and
normalized according to
the equations (6)-(7). }
\end{figure}
The fiducial model is the flat $\Lambda$CHDM model with:
$\Omega_b=0.047$,
$\Omega_c=0.233$, $\Omega_{\Lambda}=0.67$,
$\Omega_{\nu}=0.05$ ($m_{\nu}=2.1$eV),  $H_0=65$ Km s$^{-1}$ Mpc$^{-1}$,
one  relativistic  neutrino flavor, $N_{rel}=1$,
and two massive neutrino flavors, $\Gamma=0.25$, $\Gamma_{\nu}=0.021$,
$\sigma_8=0.42$, $k_{{\rm fs}}\approx 0.03$Mpc$^{-1}$ and
$k_{{\rm nl}} \approx 0.32$Mpc$^{-1}$.
We assume primordial adiabatic perturbations,
the presence of the scalar modes with the scalar spectral index
$n_s=1$, the contribution of the tensorial modes (gravitational waves)
with the spectral index $n_t=-0.09$, a
tensor-to-scalar ratio
T/S=0.1 and an optical depth to the last scattering $\tau=0$.
The independent parameters in our computation are: $\omega_b$,
 $\omega_c$, $\omega_{\nu}$, $\omega_{\Lambda}$, $n_s$, $n_t$, $T/S$,  $\tau$
and $\sigma$. We fix $\sigma_8$ at the value  given
by the equations (6)-(7) for each corresponding cosmological model and consider a
linear bias $b_I=1/\sigma_8$.

Panel a) of Figure~6 presents the CMB anisotropy power spectrum
of the fiducial model (continuous line)  and its decomposition
into scalar and tensorial contributions (the dashed lines) compared with
the CMB anisotropy experimental data from COBE, Boomerang and MAXIMA1.
Panel b) shows the polarization power spectra of the fiducial
cosmological model.
Panel c) of Figure~6 presents the matter fluctuations power spectrum
of the fiducial model and
the effective volume of the BRG sample of SDSS obtained
for the fiducial model. The solid curves are for the linear regime,
the dashed curves are obtained for the
nonlinear regime and the dash-dotted curves are obtained when both
nonlinear effects and redshift distortions are considered.
For the last case we plot the matter power spectrum
and $V_{eff}/\lambda^3$ for few choices of parameter
$\sigma$ in the equation (15).
We take for the fiducial model  $\sigma \approx 0$.
This makes the correction factor in equation (15) of order unit
for the fiducial model. \\
\begin{figure}
\plotone{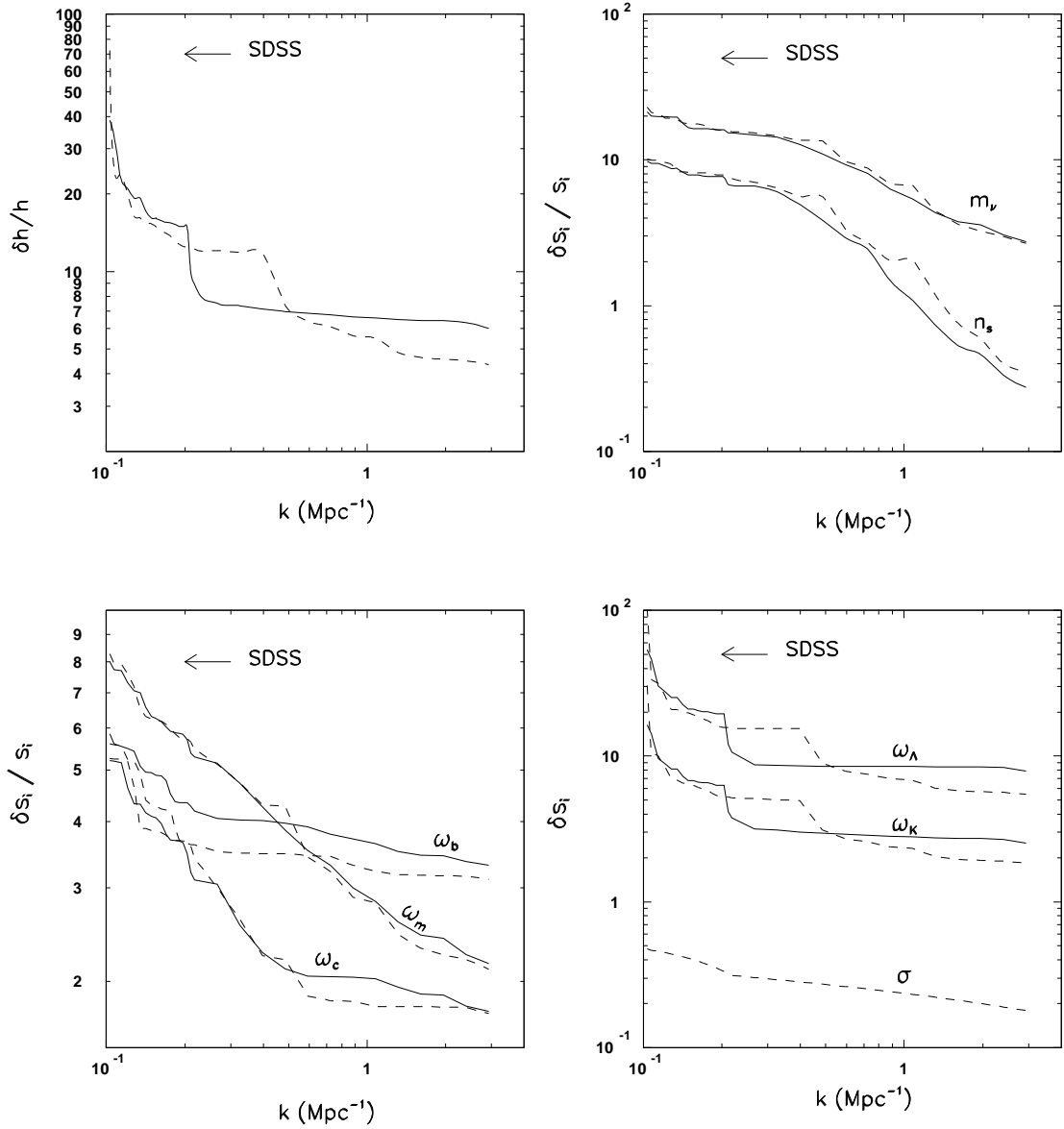}
\caption{$1-\sigma$ errors on the estimates of the cosmological
parameters that can be obtain by SDSS alone from
the linear matter power spectrum (continuous lines)
and from the matter power spectrum corrected for nonlinear effects
and redshift distortions (dashed lines) as function of $k_{max}$.
The maximum $k$ accessible to SDSS ($\sim 0.3$~Mpc$^{-1}$) is indicated by
the arrow.}
\end{figure}

Figure~7 presents the $1-\sigma$ errors on the estimates of the
cosmological parameters that can be obtained by SDSS alone
as a function of $k_{max}$.
The solid curves are obtained assuming the linear
matter power spectra in equations (19)-(21).
The dashed curves are obtained when the nonlinear effects
and the redshift distortions on the matter power spectra
are taken into account.\\
For the computation of the Fisher matrix
elements given by the equation (19) we take two-sided derivatives
of ${\cal P}(k)$ considering a step size of $\Delta s_i \approx 5\%$.\\
Parameters as $n_t$, T/S and $\tau$, not considered in Figure~7,
do not affect the matter power spectrum
for the SDSS experimental specifications considered here.
The results presented in  Figure~7
show that  the  SDSS power spectrum information alone
can give modest constraints  on most of the cosmological parameters
in both linear and nonlinear regime, strongly dependent
on the choice  of $k_{max}$ value.\\
\begin{figure}
\plotone{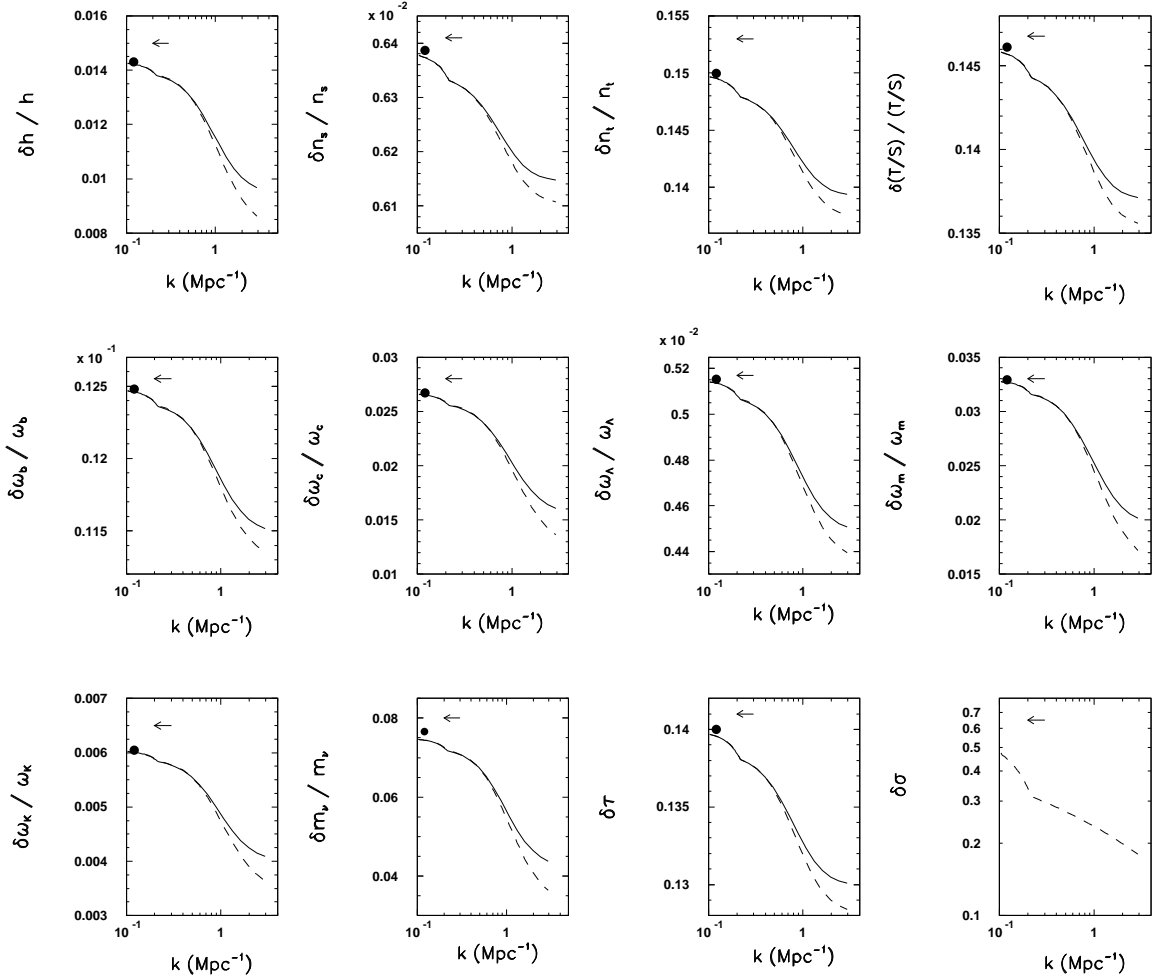}
\caption{$1-\sigma$ errors on the estimates of the cosmological
parameters from the  combined data sets of
{\sc Planck} and SDSS when:
the linear matter power spectrum is used (continuous lines),
the matter power spectrum is corrected for nonlinear effects
and redshift distortions (dashed lines).
In each case we indicate by the filled circle  the  $1-\sigma$ error
obtained by {\sc Planck} alone
from anisotropy and polarization data.
The arrow indicates the maximum $k$ accessible to the SDSS.}
\end{figure}

Figure~8 presents the $1-\sigma$ errors on the estimates of the cosmological
parameters that can be obtained by {\sc Planck} alone when the CMB anisotropy
and polarization data are taken into account and
when the combined data sets (CMB anisotropy and
polarization and matter density fluctuations) are taken into account
for different values of $k_{max}$.
We consider for the {\sc Planck} surveyor a sky coverage  $f_{sky}=0.85$.
Figure~8 shows that,
although the CMB anisotropy and polarization
measurements tend to dominate the constraints on most of the cosmological
parameters, the additional small scale
LSS data  ($k > 0.1$ Mpc$^{-1}$)
help to break the parameter degeneracies.

\section{Conclusions}

In this paper we study the complementarity between the
cosmological information obtainable with the {\sc Planck}
surveyour and the large scale structure redshift surveys in
$\Lambda$CHDM cosmologies.

We compute the initial full phase-space neutrino distribution
function  for CHDM and $\Lambda$CHDM models by using numerical simulations.
We start from the HDM density fluctuation
power spectrum in the normalization indicated by the analysis of
the local cluster X-ray
temperature function
and derive the initial neutrino phase-space distribution
at each spatial wave number $k$ by using
Zel'dovich approximation.
The neutrino phase-space distributions obtained in this way
are  implemented in the CMBFAST code
as initial neutrino momentum distributions for the
integration of the coupled linearized Einstein, Boltzmann and fluid
equations in $k$-space and the computation of CMB power spectra and matter
transfer functions.

We find that the relative bias between the CMB anisotropy
power spectrum and the matter density fluctuations power spectrum in
the COBE/DMR normalization is given by the CDM component normalized
to the abundancy of the rich clusters at the present time.

Taking into account the redshift distortions and
nonlinear evolutionary effects on the matter density fluctuations
power spectrum, we constrain an 11-dimensional parametrization of the
$\Lambda$CHDM model, when the combined CMB and LSS data are taken into account
in the {\sc Planck} and BRG sample of SDSS experimental specifications.
We find that, depending on the maximum spatial wave number,
the combined CMB and LSS data can better constrain most of the cosmological
parameters.



\acknowledgments

It is a pleasure to thank U.~Seljak and M.~Zaldarriaga
for the use of the CMBFAST code (v3.2) employed
in the computation of the CMB power spectra and of
the matter transfer functions.




\end{document}